\documentclass[12pt]{article}

\usepackage[english]{babel}

\usepackage[letterpaper,top=2cm,bottom=2cm,left=3cm,right=3cm,marginparwidth=1.75cm]{geometry}

\usepackage[
    backend=biber,
    style=authoryear, 
    maxbibnames=3,    
    minbibnames=1,    
    dashed=false,     
    mincitenames=1,
    maxcitenames=1,
]{biblatex}
\usepackage{amsmath}
\usepackage{graphicx}
\usepackage[colorlinks=true, allcolors=blue]{hyperref}
\usepackage{authblk}
\usepackage{multirow, booktabs}
\usepackage{amsmath, amssymb, bbold, bm}
\usepackage{graphicx}
\usepackage{subcaption}
\usepackage{tikz}
\usetikzlibrary{arrows.meta, positioning, fit, backgrounds}

\title{Improving Eddy Detection from Wide-Swath Altimetry with Multimodal Data-Driven Models}
\author[1]{M.~Broccoli}
\author[1]{A.~Cipollone}
\author[2]{A.~Bonaduce}
\author[3]{E.~Carli}
\author[4]{F.~Le~Guillou}
\affil[1]{CMCC Foundation -- Euro-Mediterranean Center on Climate Change, Bologna, Italy}
\affil[2]{Nansen Environmental and Remote Sensing Center (NERSC) and Bjerknes Center for Climate Research, Bergen, Norway}
\affil[3]{European Space Agency (ESA) -- ESRIN, Frascati, Italy}
\affil[4]{Datlas, Grenoble, France}

\date{\today}

\begin{document}
\maketitle

\begin{abstract}
Physically-based eddy detection algorithms usually rely on hard-coded thresholds and constraints that make the detected structures sensitive to small variations in the underlying fields.
One way to relax such conditions is the use of machine-learning methodologies that transform the full problem into an optimization process. 
We present a multimodal, multi-task, data-driven eddy detection framework based on a shallow convolutional U-Net, trained to jointly segment eddy structures and reconstruct the co-located spatial sea surface temperature (SST) anomaly using a dynamically-based reconstruction as target.
Two model variants are developed: NADIRnet, trained on conventional nadir altimetry, and SWOTnet, trained through a curriculum that incorporates SWOT-enriched maps.
Both models achieve high pixel- and structure-wise detection scores against the ground-truth, with most residual disagreement attributable to boundary placement rather than missed or spurious detections.
The coupled SST regression task, enforced through boundary-coupling and gradient-matching loss terms, is shown to sharpen and regularise the detected eddy boundaries.
Compared to the dynamically-based reference, both models recover a substantially larger and finer eddy population;
notably, SWOTnet's higher-resolution training signal transfers to conventional nadir altimetry thanks to the higher-resolution SST channel, detecting more eddies than a nadir-trained model on the same data.
The framework further generalises, even without retraining, to altimetry products and processing levels unseen during training, including SWOT L3 wide-swath retrievals and an independent gridded product.
Together with its low computational cost, this framework is thus promising for near-real-time operational deployment and further downstream applications.
\end{abstract}

\section{Introduction}
Satellite altimetry and related processed products have been fundamental to monitor and understand ocean variability for more than three decades \parencite{ traon_chapter_2001, meyssignac_how_2023}.
These products have been widely used to detect the coherent part of ocean turbulence and quantifying its anomalous information content as well as the corresponding feedbacks to the ocean circulation \parencite{zhang_oceanic_2014,zhao_meridional_2018}.
Processes associated to coherent vortices in particular have been the focus of  a variety of studies ranging from eddy polarity to anomalous tracer transport~\parencite{gaube_regional_2014,dong_global_2014,raj_quantifying_2016, raj_interaction_2020}.
While the temporal sampling of gridded L4 product has reached a stable daily frequency already a decade ago, the recent wide-swath altimetry concept --- delivered by the Surface Water and Ocean Topography (SWOT) mission --- is pushing towards future altimetry products with an unprecedented spatial resolution thanks to its two-dimensional mapping and sampling of the ocean surface~\parencite{fu_high-resolution_2004,morrow_global_2019}.
While wide-swath observations directly resolve 2D sea-surface topography down to about $15\text{--}30\, \text{km}$ in along-swath products~\parencite{fu_wide-swath_2026}, their integration into multi-mission L4 gridded maps yields an effective resolution gain of approximately $5\text{--}10\,\text{km}$ globally (locally up to $20\,\text{km}$ in energetic regions;~\cite{ballarotta_integrating_2025}).
Combining advances in mapping methodologies with future wide-swath constellations thus provides valuable opportunities to better constrain fine-scale mesoscale variability~\parencite{traon_use_2015, ballarotta_integrating_2025}.

Despite these resolution gains, present satellite altimetry network is not able to sample the global ocean with uniform accuracy across all basins, therefore gap-free L4 altimetry products heavily rely on Optimal Interpolation weighting.
In this context, the identification of real eddies is not straightforward and a  set of necessary physical conditions is generally used in literature for the detection, although such set is not uniquely identified \parencite{nencioli_vector_2010, chelton_global_2011, souza_comparison_2011, mason_new_2014,faghmous_daily_2015, ashkezari_oceanic_2016}.
As surface dynamically-based eddy-detection methods generally look for extremes in altimetry maps to detect eddies and remove spurious patterns~\parencite{chelton_global_2011, mason_new_2014}, it has been shown with a 3D analysis that these methods still include patterns that do not penetrate the subsurface layers~\parencite{cipollone_benchmarking_2017}.

To mitigate this problem, \cite{bonaduce_2026} proposed combining SST and altimetry maps to identify ``active'' eddies, i.e.~elevation or depression patterns in the sea level field that express a significant internal temperature anomaly relative to the surrounding waters, indicative of dynamical or thermodynamic processes.
The active eddy populations were evaluated in terms of their induced SST and heat transport anomalies, allowing a direct comparison between structures detected in conventional altimetry and those resolved in SWOT-enriched fields.
In particular, the enhanced sampling capability provided by SWOT markedly increases the fraction of eddy kinetic energy associated with coherent structures, revealing a substantial component of the mesoscale circulation that is only partially resolved by conventional altimetry.

In parallel to dynamical eddy-detection algorithm, data-driven eddy detection methods have been growing in importance year by year, with convolutional neural network (CNN) architecture competing in efficiency with dynamically-based eddy detection schemes~\parencite{xu_application_2021}.
The former, in fact, are able to combine information from different sources and scales, for example fine-scale infrared SST structures that are not resolved in gridded altimetry~\parencite{moschos_computer_2023}, or ocean-color chlorophyll-a~\parencite{liu_dual-branch_2023} or SST~\parencite{liu_detection_2023}, which can improve detection where altimetry alone is noisy or sparsely sampled.
Starting from initial attempts with simple, encouraging machine learning models~\parencite{ashkezari_oceanic_2016}, the complexity of designed network has largely grown in the last decade, together with the risk of over-fitting and losing the small-scale details~\parencite{santana_oceanic_2022, huo_high_2024}.

While earlier convolutional approaches classify pixels of altimetry maps mostly over limited regions~\parencite{lguensat_eddynet_2018,franz_ocean_2018,duo_oceanic_2019, xu_oceanic_2019, xu_application_2021}, more recent studies fuse altimetry with SST maps at global scale~\parencite{liu_detection_2023} or limited to the ice-free ocean between $60^\circ\,\text{S}$ -- $60^\circ\,\text{N}$~\parencite{qiu_phyeddy-net_2026}.
Nonetheless, all of the above algorithms have been trained on conventional nadir-based products, and detection on SWOT data has so far been performed with dynamical algorithms, applied either to merged L4 products~\parencite{verger-miralles_swot_2025,gomez-navarro_new_2026} or directly to the swaths~\parencite{zhang_automated_2025,de_marez_high-latitude_2026,fu_wide-swath_2026}, so that the question of how a data-driven model trained on wide-swath-enriched fields performs on conventional altimetry has not been addressed, to our knowledge.
Moreover, most data-driven algorithms are evaluated with pixel-wise scores against the dynamical reference, which is not necessarily a faithful representation of the eddy field.
Indeed, gridded altimetry can distort eddy size and intensity, especially at high latitudes~\parencite{amores_up_2018, de_marez_high-latitude_2026}.

In this work, we present a new data-driven eddy detection algorithm consisting of a shallow CNN model based on the U-Net architecture~\parencite{ronneberger_u-net_2015}, that exploit the knowledge of the eddy population obtained in~\cite{bonaduce_2026}.
The network is formulated as a multi-modal, multi-task model, that receives altimetry and SST maps as input and simultaneously performs binary eddy segmentation and reconstructs the co-located spatial SST anomaly.
While~\cite{bonaduce_2026} retains only thermodynamically ``active'' eddies, here our detection targets the population consisting of both active and inactive eddies, using the coupled SST reconstruction branch to physically regularise and sharpen the detected boundaries.
By developing two model variants --- one trained on conventional nadir altimetry (NADIRnet) and one fine-tuned on SWOT-enriched fields (SWOTnet) --- we assess whether high-resolution spatial features learned from SWOT can transfer to conventional altimetry, opening new avenues for reprocessing the multi-decadal altimetry record since 1993.

The framework presented here differs from previous work in that: (i)~it processes the full global $1/10^\circ$~grid in a single forward pass;
(ii)~it is trained on, and applied to, SWOT-enriched altimetry through a curriculum strategy, and transfers to conventional altimetry;
(iii)~SST is not only an input but is reconstructed jointly with the eddy mask, through boundary-coupling and gradient-matching loss terms that regularise the detected contours;
(iv)~detection is evaluated at both pixel and structure level;
(v)~it generalises, without retraining, to other products unseen during training and to wide-swath L3 data.
Weights and code are openly accessible~\parencite{broccoli_water-ml-eddy-detection_2026}.

The remainder of the paper is structured as follows.
Section~\ref{sec:data} collects and describes the altimetry and SST data that are employed for training, while Sec.~\ref{sec:model} presents the architecture of the model, its loss function and training configurations.
In Sec.~\ref{sec:evaluation} the evaluation protocol is described, distinguishing in particular between pixel-wise and structure-wise evaluations for the eddy detection, that are both carried out.
Section~\ref{sec:results} presents the main results, including the results of the evaluation on the validation data and the comparisons against the target data, showing the improvements of the ML-based detection on the dynamical detection; further possible applications are also discussed.
We then conclude in Sec.~\ref{sec:conclusions}.

\section{Data}\label{sec:data}
In this section we describe the datasets that are employed for training and evaluation of the ML-based detection algorithm.
For consistency reasons, all data is gridded to the same global $1/10^\circ$~resolution regular grid.

\paragraph{Conventional altimetry}
Global ocean gridded L4 Sea Surface Heights (SSH) and derived variables generated by the DUACS multimission altimeter data processing system are considered as conventional altimetry maps~\parencite{cmems_conventional_altimetry}.
The Sea Level Anomaly (SLA) is estimated by Optimal Interpolation, merging the L3 along-track measurement from different altimeter missions, including Jason-3, Sentinel-3A, HY-2A, Saral/AltiKa, Cryosat-2, Jason-2, Jason-1, T/P, ENVISAT, GFO and ERS1/2.
SLA is computed with respect to a twenty-year (1993--2012) mean, from which geostrophic velocity anomalies are derived.
We downloaded data from the year 2022 to 2024.

\paragraph{SWOT-enriched altimetry}
Wide-swath altimetry data from the SWOT mission are employed to train the network on a higher-resolution field compared to conventional altimetry.
SWOT L4 products\footnote{The data is processed by SSALTO/DUACS and distributed by AVISO (\url{https://www.aviso.altimetry.fr}) supported by CNES. DOI:~\href{https://doi.org/10.24400/527896/A01-2024.007}{10.24400/527896/A01-2024.007}.} are processed with a Multiscale Inversion of Ocean Surface Topography (MIOST) method~\parencite{ubelmann_reconstructing_2021} applied to both nadir and wide-swath altimetry tracks.
Compared to nadir-only maps, the integration of SWOT into MIOST yields an effective spatial resolution enhancement of $5\text{-}10\,\text{km}$ on global average --- locally up to $15\text{-}20\,\text{km}$ in energetic boundary currents --- and reduces SSH mapping error by approximately $10\%$ in highly dynamic regions~\parencite{ballarotta_integrating_2025}.
The resulting gap-free product is distributed as daily fields on a regular $1/10^\circ$~horizontal grid.
At the time of this study, available data covered the period from 27 Jul 2023 to 1 May 2024, which were downloaded.

\paragraph{SST field}
The ESA Climate Change Initiatives (CCI) SST products map the temperature of the global oceans using observations from multiple infrared satellite sensors, including the Advanced Very High Resolution Radiometer (AVHRR), Advanced Along-Track Scanning Radiometer (AATSR) and Sea and Land Surface Temperature Radiometer (SLSTR) series.
The data are distributed at different processing levels.
In this study we exploit the ESA CCI SST L4 data~\parencite{good_esa_2024}, which consists of gap-free daily mean SST analyses at $1/20^\circ$ resolution.
Data from the year 2022 to 2024 were downloaded; at the time of the study the year 2024 was covered until June~22.

\section{Model}\label{sec:model}
The data-driven eddy detection system is a multi-modal and multi-task CNN based on the U-Net architecture~\parencite{ronneberger_u-net_2015}, implemented in TensorFlow~\parencite{tensorflow2015-whitepaper}.
The model jointly performs a binary segmentation of eddy structures and a regression of the spatial SST anomaly field as defined in~\cite{bonaduce_2026}.
In this section the model's detailed architecture is described, together with its loss function and training strategies.
The model weights and relevant code are available through~\cite{broccoli_water-ml-eddy-detection_2026}.

\subsection{Architecture}\label{sec:architecture}
The network input is of shape $(H, W, C_\textrm{input})$, with $H=1701$ and $W=3600$ corresponding to the global $(1701\times 3600)$ grid of the SWOT L4 MIOST altimetry product, and with $C_\textrm{input} = 6$ physical-variables channels stacked together.
These input variables correspond to SSH, SST, zonal and meridional components of the geostrophic velocity $(u, v)$, and surface phase velocities $(\Phi_1, \Phi_2)$ derived from the geostrophic velocity components as
\begin{equation}
    \Phi_1 = \operatorname{atan2}(v, u) \cdot \frac{180}{\pi} \,, \qquad \Phi_2 = \Phi_1 \pmod{360} \,.
\end{equation}
The inclusion of phase velocities among the inputs follows~\cite{ashkezari_oceanic_2016}, where it is shown that eddy core emerges as circular patterns of phase values producing a distinct signature that can be detected by ML algorithms.
Each input channel is normalised according to its maximum absolute value, so that physical zeros are zeros also in the model space; the SST (temporal) anomaly over the training dataset is further computed before normalisation and used for the SST channel.

The core of the network is composed of an encoder and a decoder.
The encoder consists of $3$~downsampling blocks, each of which applies a 2D~convolution with kernel size~$(4\times4)$ and strides~$2$, ReLU activation and batch normalisation.
Each downsampling stage halves the spatial resolution while doubling the channel width, which is set to~$64$ in the first convolutional layer.
The decoder mirrors the encoder, although using transposed 2D~convolution so that the channel width is halved while spatial resolution is doubled.
Dropout layers are included for regularization, as well as skip connection between encoder and decoder blocks following the original U-Net architecture.
The encoder-decoder design with convolutional layers lets the network preserve fine spatial details while still aggregating features across multiple scales.
This design has already been proven beneficial for ocean applications over more spatial-agnostic methods~\parencite{broccoli_assimilation_2025}.

A transposed convolution layer maps the last decoder features to the output of the model, which is of shape~$(H,W,C_\textrm{output})$ with $C_\textrm{output} =2$.
The first output channel is the eddy-mask logit~$\hat{\bm z}$, while the second channel is the network reconstruction of the spatial SST anomaly~$\hat{\bm s}$.
We stress that the eddy-detection mask is raw logit at this stage, so that a sigmoid function $\sigma$ has to be applied to the network output to convert it into proper eddy-detection mask; we will discuss this into more details in Sec.~\ref{sec:evaluation}.
A summary of the model's inputs and outputs is in Tab.~\ref{tab:channels}.

\begin{table}[h]
    \centering
    \begin{tabular}{@{}cclp{6cm}@{}}
        \toprule
        & {$C$} & {Variable} & {Description} \\
        \midrule
         \multirow{6}{*}{{input}} & 1  & SSH             & Sea surface height \\
         & 2                          & SST             & Sea surface temperature \\
        & 3                          & $u$             & Zonal geostrophic velocity \\
         & 4                         & $v$             & Meridional geostrophic velocity \\
         & 5                         & $\Phi_1$        & Surface phase velocity \\
         & 6                         & $\Phi_2$        & Surface phase velocity \\
        \midrule
         \multirow{2}{*}{{output}} & 1 & $\hat{\bm z}$ & Eddy mask in logit \\
         & 2                         & $\hat{\bm s}$   & Spatial SST anomaly \\
        \bottomrule
    \end{tabular}
    \caption{Model's input and output channels. All fields have spatial dimension~$1701\times 3600$.}
    \label{tab:channels}
\end{table}

\subsection{Loss Function}
Since the network is trained for two coupled tasks, the loss function is a (weighted) sum of several terms, each targeting a different aspect of the desired output.
Let $Y = (\bm{m}, \bm{s})$ denote the two-channel target, with $ \bm{m} \in \{0,1\}^{H \times W}$ the ground-truth binary eddy mask and $\bm{s} \in \mathbb{R}^{H\times W}$ the normalized SST anomaly field.
Let $\hat{Y} = (\hat{\bm{z}}, \hat{\bm s})$ denote the network output, with $\hat{\bm{z}} \in \mathbb{R}^{H \times W}$ the raw mask logit and $\hat{\bm s} \in \mathbb{R}^{H \times W}$ the predicted SST anomaly, and let $\hat{\bm m} = \sigma(\hat{\bm{z}})$ be the sigmoid-activated mask probability.
$N$ denotes the number of valid pixels in a batch.

\paragraph{Detection term}
This term is a weighted binary cross-entropy loss computed directly from logits:
\begin{equation}
\mathcal{L}_{BCE} = \frac{1}{N} \sum_{i,j} w_{ij}\Big[-m_{ij}\log\sigma(\hat z_{ij}) - \big(1-m_{ij}\big)\log\big(1-\sigma(\hat z_{ij})\big)\Big]\,,
\label{eq:bce}
\end{equation}
where the per-pixel weight
\begin{equation}
w_{ij} = {\big[\lambda_e m_{ij} + \lambda_o (1-m_{ij})\big]} \cdot {\mathbb{1}_{ij}}
\label{eq:weight}
\end{equation}
compensates for the class imbalance intrinsic to eddy segmentation giving weight $\lambda_e = 2$ to eddy pixels and weight $\lambda_o = 0.67$ to non-eddy pixels, and zeroes out land (or invalid-data) pixels with $\mathbb{1}_{ij}$.

\paragraph{Regression term}
This term is a mean squared error (MSE) to train the reconstruction of the spatial SST anomaly co-located with the eddy:
\begin{equation}
\mathcal{L}_{MSE} = \frac{1}{N} \sum_{ij}\big(s_{ij} - \hat{s}_{ij}\big)^2 \, .
\label{eq:mse}
\end{equation}

\paragraph{Boundary--SST coupling term}
To explicitly couple the two output branches, the boundary $\bm{B}(\cdot)$ of the eddies is extracted from a given mask and combined with the spatial SST anomaly signal.
In particular, this loss term penalises disagreement between the spatial SST signal restricted to the eddy border in the target and predicted fields, together with a boundary-alignment penalty between the two boundaries themselves as:
\begin{equation}
\mathcal{L}_{coupling} = \underbrace{\frac{1}{N}\sum_{i,j}\Big(B(\bm m)_{ij}\, s_{ij} - B(\hat{\bm m})_{ij}\, \hat s_{ij}\Big)^2}_{\text{boundary--SST product MSE}} + \lambda_b \underbrace{\frac{1}{N}\sum_{i,j}\Big(B(\bm m)_{ij} - B(\hat{\bm m})_{ij}\Big)^2}_{\text{boundary alignment}} \, .
\label{eq:coupling}
\end{equation}
The weight $\lambda_b = 0.1$ keeps the order of magnitude of the two terms on a comparable scale.

\paragraph{Gradient term}
To discourage the blurry reconstructions typical of pixel-wise regression losses such as MSE, a gradient-matching term compares the horizontal and vertical gradients ($\nabla_x$ and $\nabla_y$, respectively) of the true and predicted spatial SST anomaly fields,
\begin{equation}
\mathcal{L}_{gradient} = \frac{1}{N}\sum_{i,j}\Big(\big|(\nabla_x \hat{\bm s})_{ij} - (\nabla_x \bm s)_{ij}\big| + \big|(\nabla_y \hat{\bm s})_{ij} - (\nabla_y \bm s)_{ij}\big|\Big) \, ,
\label{eq:gradient}
\end{equation}
encouraging sharp and physically plausible spatial SST anomaly boundaries in the output.

\paragraph{Total loss} The four terms are summed with a fixed $\lambda=1000$ weight
chosen to balance their differing magnitudes:
\begin{equation}
\mathcal{L}_{total} = \mathcal{L}_{BCE} + \lambda(\mathcal{L}_{MSE} + \mathcal{L}_{coupling}) + \mathcal{L}_{gradient} \, .
\label{eq:total}
\end{equation}
The weight on the MSE and coupling terms reflect the fact that both operate on small-magnitude normalized SST anomalies, whose squared errors are numerically much smaller than the cross-entropy term unless upweighted; this keeps the gradients of all four terms on a comparable scale during backpropagation.

\subsection{Training}
\begin{figure}[htbp]
\centering
\includegraphics[width=\textwidth]{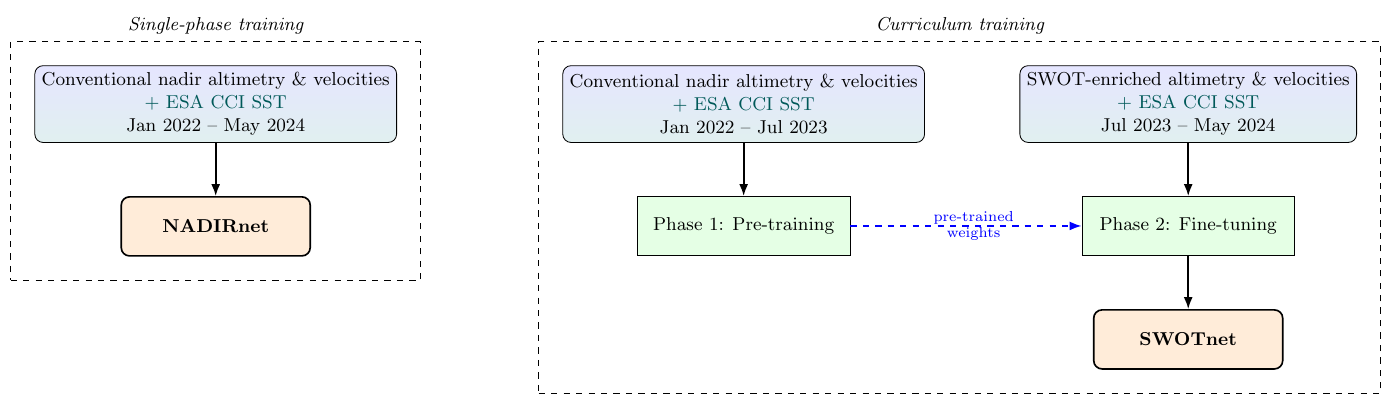}
\caption{Training scheme for the two model variants. NADIRnet is trained in a single phase on conventional nadir altimetry data. SWOTnet undergoes a two-phase curriculum: a pre-training phase on conventional altimetry, followed by a fine-tuning phase on SWOT-enriched altimetry initialized from the pre-trained weights. ESA CCI SST maps are common to both.}
\label{fig:training_scheme}
\end{figure}

The architecture described in Sec.~\ref{sec:architecture} is used to develop two distinct models, i.e.~NADIRnet, trained on conventional nadir altimetry, and SWOTnet, which incorporates SWOT-enriched data:
\begin{itemize}
    \item NADIRnet is trained on conventional altimetry maps and associated velocities from January 2022 to May 2024; training is conducted with a learning rate (LR) of $1\times 10^{-4}$.
    \item SWOTnet undergoes a two-phase curriculum training due to the limited temporal coverage of SWOT-enriched data; in the first phase, the model is pre-trained on conventional altimetry (January 2022--July 2023) at $\text{LR} = 1\times 10^{-4}$, while in the second phase the model is fine-tuned on SWOT-enriched data (July 2023--May 2024) with a reduced learning rate of $\text{LR} = 1\times 10^{-5}$ to avoid catastrophic forgetting.
\end{itemize}
Both models exploit the ESA CCI L4 product for SST, and are trained on global data at 1/10$^\circ$ resolution with Adam optimizer.
Ground-truth data for eddy detection and spatial SST anomaly reconstruction are derived from~\cite{bonaduce_2026}.
Across all training phases, data is partitioned into an 80/20\% training and validation split.
The training schemes are summarised in Fig.~\ref{fig:training_scheme}.
Training of the models took about~$9\,\mathrm{h}$ for the single-phase curriculum on a single NVIDIA~A100~GPU, and similarly for the two-phase curriculum with about $8\,\mathrm{h}$ for the pre-training and $1\,\mathrm{h}$ for the fine-tuning phases.
On the same hardware, prediction on a single-day global map takes about $0.6\,\mathrm{s}$.

\section{Evaluation}\label{sec:evaluation}
The models' dual-task nature, i.e.~eddy detection and spatial SST anomaly reconstruction, requires distinct evaluation protocols.
Evaluation is performed on the validation split of the data.

\subsection{Eddy detection evaluation}
Detection performance is assessed at two levels of granularity.  
Pixel-wise scores are the standard in the eddy-detection literature~\parencite{lguensat_eddynet_2018, duo_oceanic_2019, xu_application_2021, santana_oceanic_2022} and are close to the training objective, but they are insensitive to whether individual structures are correctly identified, merged or split.
We therefore complement them with a structure-wise evaluation, adapted from object-detection benchmarks, in which each predicted eddy is matched to a reference eddy through Intersection-over-Union~\parencite{rainio_evaluation_2024};
it is more physics-oriented, since eddies are the objects of interest, and stricter, since a hit requires agreement of the whole structure rather than of its core.
For both levels, we report Precision, Recall and the F\textsubscript{1} scores, that are common $[0,1]$-bounded metrics for binary classification tasks, where higher values are desirable~\parencite{rainio_evaluation_2024}.
These metrics are defined based on True Positives (TP) or hits, False Positives (FP) or false alarms, and False Negatives (FN) or misses:
\begin{equation}\label{eq:eval_scores}
    \mathrm{Precision} = \frac{\mathrm{TP}}{\mathrm{TP} + \mathrm{FP}} \,, \quad
    \mathrm{Recall} = \frac{\mathrm{TP}}{\mathrm{TP} + \mathrm{FN}} \,, \quad
    \mathrm{F}_1 = 2\,\frac{\mathrm{Precision} \cdot \mathrm{Recall}}{\mathrm{Precision} + \mathrm{Recall}} \,.
\end{equation}

Since the models output the eddy mask as raw logits, an operating threshold must first be selected before a binary mask can be produced and scored.
This is done on the first 30 days of the validation period by sweeping 25 candidate logit thresholds linearly spaced over $[-2.5, 2.5]$, binarizing the predicted mask at each threshold, and computing the scores in~\eqref{eq:eval_scores}.
The threshold maximizing the F$_1$ score is selected for final evaluation.
Additionally, predicted structures with equivalent radius\footnote{Radius of a disk with exactly the
same area.} smaller than 10 km are removed before matching in order to suppress spurious sub-grid detections that would otherwise inflate false-alarm counts~\parencite{bonaduce_2026}.

For structure-wise evaluation, each predicted eddy is matched to the reference eddy with which it has maximum Intersection-over-Union (IoU),
and classified as a \emph{hit} if $\mathrm{IoU} \geq 0.5$, a \emph{partial match} if $0 < \mathrm{IoU} < 0.5$, a \emph{false alarm} if $\mathrm{IoU} = 0$, or a \emph{miss} if a reference eddy has no overlapping prediction.
To provide a comprehensive view of detection quality, two summary scores are reported for the structure-wise evaluation: a \emph{strict} score, in which only $\mathrm{IoU} \geq 0.5$ counts as a true positive and partial matches count as false negatives, and a \emph{soft} score, in which partial matches also count as true positives.
Comparing the two indicates how much of the network's apparent performance comes from near-miss detections rather than tight boundary agreement.
Table~\ref{tab:detection_scores} reports precision, recall, and F$_1$ for both models at both granularities, with structure-wise scores given as $[\text{strict}, \text{soft}]$~ranges.
Table~\ref{tab:structure_breakdown} further decomposes the structure-wise matching into the fraction of hit, partial match, miss, and false alarm structures, normalized by the total number of reference structures.

Both levels measure agreement with the dynamical reference, not with the true eddy field.
Perfect agreement is neither expected nor desirable, as the reference is the output of a sequence of thresholds and constraints that can merge nearby vortices or retain surface-only patterns~\parencite{cipollone_benchmarking_2017}, and a model reproducing it exactly would inherit these limitations.
We therefore regard a high but sub-unity score as the appropriate target, with residual disagreement concentrated at boundaries and merged structures (see Sec.~\ref{sec:results_validation}).

\begin{table}[htbp]
    \centering
    \begin{tabular}{@{}lcccccccc@{}}
        \toprule
        & & \multicolumn{3}{c}{Pixel-wise} & \multicolumn{3}{c}{Structure-wise} \\
        \cmidrule(lr){3-5} \cmidrule(lr){6-8}
        Model & Logit thr. & Precision & Recall & F$_1$ & Precision & Recall & F$_1$ \\
        \midrule
        NADIRnet & 0.417 & 0.87 & 0.75 & 0.81 & [0.82, 0.85] & [0.75, 0.92] & [0.77, 0.88] \\
        SWOTnet  & 0.208 & 0.85 & 0.73 & 0.78 & [0.78, 0.82] & [0.69, 0.91] & [0.73, 0.86] \\
        \bottomrule
    \end{tabular}
    \caption{Pixel-wise and structure-wise detection scores for the two models at their
    respective optimal logit thresholds. Structure-wise values are reported as
    $[\text{strict}, \text{soft}]$ ranges, where the lower bound counts partial matches
    as false negatives and the upper bound counts them as true positives.}\label{tab:detection_scores}
\end{table}

\begin{table}[htbp]
    \centering
    \begin{tabular}{@{}lccccc@{}}
        \toprule
        Model & Logit thr. & Hit & Partial match & Miss & False alarm \\
        \midrule
        NADIRnet & 0.417 & 0.73 & 0.19 & 0.08 & 0.16 \\
        SWOTnet  & 0.208 & 0.69 & 0.22 & 0.09 & 0.19 \\
        \bottomrule
    \end{tabular}
    \caption{Structure-wise matching outcomes, normalized by the total number of
    reference structures, at the optimal logit threshold.}\label{tab:structure_breakdown}
\end{table}

\subsection{Regression evaluation}
The spatial SST anomaly reconstruction output of the models is evaluated independently. 
Root mean squared error (RMSE) maps are computed pixel-wise across the validation period against the target maps.
Qualitative side-by-side maps of ground-truth, prediction, and error residuals are produced at sampled validation days to visually assess whether the reconstructed anomaly correctly localizes around detected eddy structures, as a direct qualitative check on the boundary-SST coupling behaviour the loss function is designed to encourage.
Some examples are reported in the next section.

\section{Results}\label{sec:results}
In this section we present the results of the ML-based eddy-detection models.
We will present some visualisations of the outputs, in particular against the target data, and discuss the impact of the different terms of the loss function to the final detection.
Differences in the eddy populations detected by the two NADIRnet and SWOTnet models will be outlined, arguing for the benefit of a model trained on higher resolution data even when applied to lower resolution altimetry.
Finally, possible future applications of the data-driven detection are presented, highlighting its robustness and versatility over different datasets not seen during training or at different processing levels, e.g.~L3 satellite altimetry retrievals.

\subsection{Results on validation data}\label{sec:results_validation}
Table~\ref{tab:detection_scores} reports the pixel-wise and structure-wise detection scores for NADIRnet and SWOTnet at their respective optimal logit thresholds.
At the pixel level, NADIRnet slightly outperforms SWOTnet, and a similar pattern holds at the structure level, where NADIRnet's strict/soft F\textsubscript{1} range is marginally higher than SWOTnet's range.
This can be explained since NADIRnet is trained and evaluated on the smoother, lower-resolution conventional altimetry, on which the definition of eddy boundaries against the target is comparatively less ambiguous, whereas SWOTnet must resolve finer-scale structures present in the higher-resolution SWOT-enriched maps, increasing the possibility of boundary disagreements.
The gap between the strict and soft structure-wise scores is comparable for both models, suggesting that a comparable fraction of each model's apparent underperformance stems from boundary displacements rather than from missed or spuriously detected eddies.
This is corroborated by the outcome breakdown in Table~\ref{tab:structure_breakdown}, as for both models the hit fraction dominates, with partial matches the second most frequent outcome and true misses --- less than 10\% --- the rarest, confirming that the models rarely fail to detect an eddy altogether but more often disagree with the target on its precise extent.

\begin{figure}[htbp]
    \centering
    \includegraphics[width=0.75\linewidth]{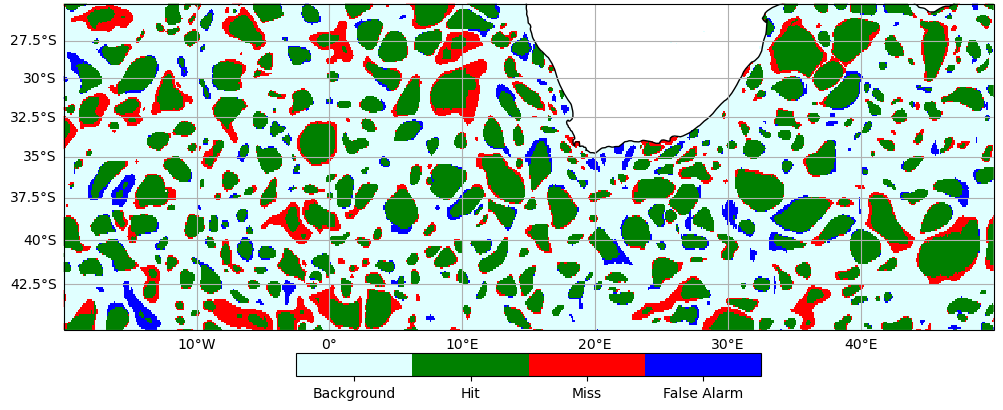}
    \caption{Pixel-wise comparison between SWOTnet and target detections zoomed-in over the Agulhas region on 2024-03-27. Green: eddy in both (hit); red: eddy in target only (miss); dark blue: eddy in SWOTnet only (false alarm); light blue: background in both.}
    \label{fig:eddy_pixels}
\end{figure}

\begin{figure}[htbp]
    \centering
    \begin{subfigure}{0.75\textwidth}
        \centering
        \includegraphics[width=\linewidth]{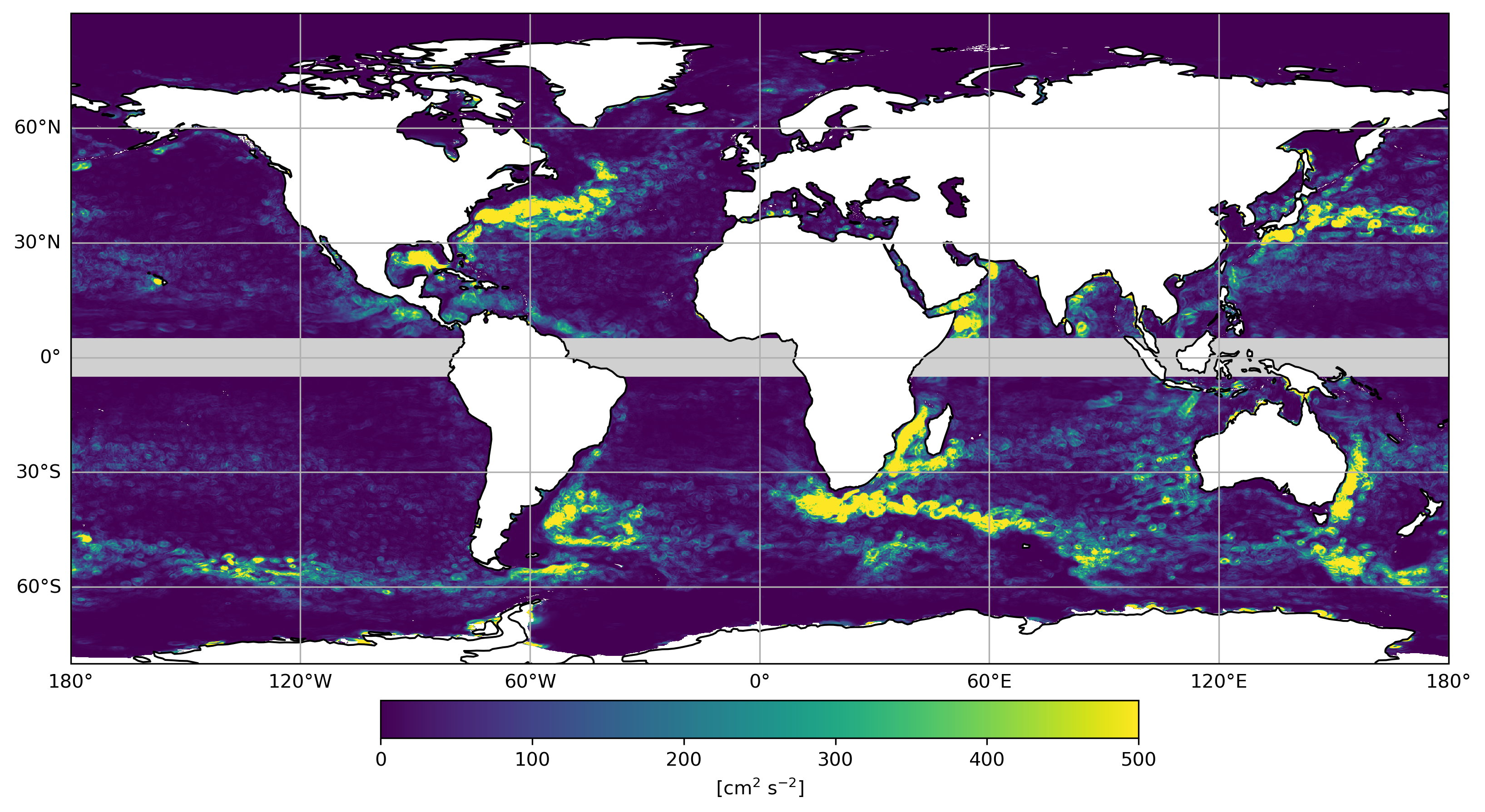}
        \caption{EKE SWOTnet}
        \label{fig:eke_swotnet}
    \end{subfigure}
    \vspace{1em}
    \begin{subfigure}{0.75\textwidth}
        \centering
        \includegraphics[width=\linewidth]{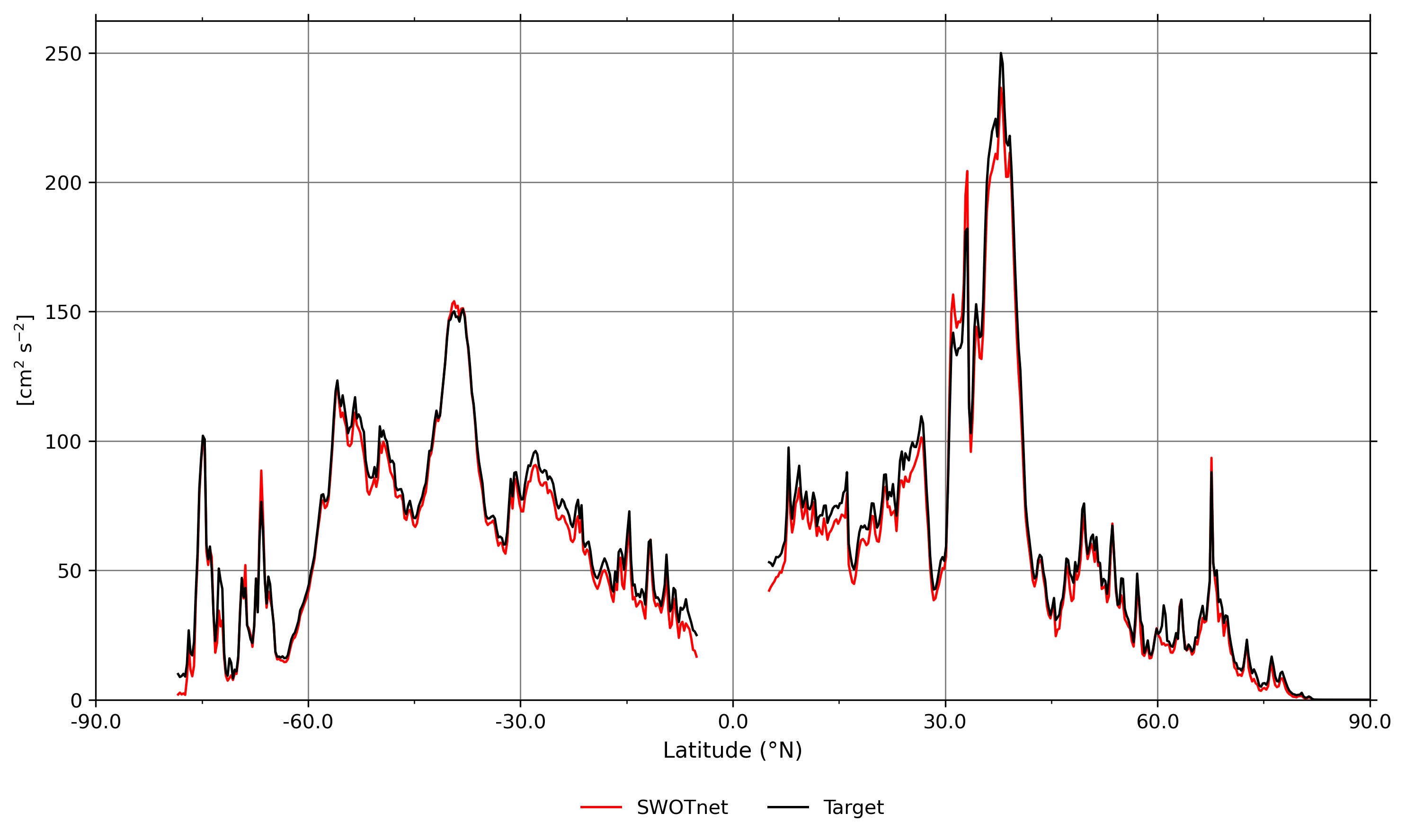}
        \caption{EKE zonal average}
        \label{fig:eke_diff}
    \end{subfigure}

    \caption{Eddy kinetic energy induced by the eddies detected by SWOTnet~\subref{fig:eke_swotnet} over the whole training and validation period. Comparison with the EKE resulting from the target eddies is in~\subref{fig:eke_diff}, where the zonal average of the two is displayed.}
    \label{fig:eke}
\end{figure}

A direct inspection also confirms that the ML-based models are able to detect the target's structure cores in most cases.
Figure~\ref{fig:eddy_pixels} shows a pixel-wise comparison between SWOTnet and the target detection over the Agulhas region, chosen as a region of high mesoscale variability.
Overall, the target pixels are correctly classified (green) in the eddy cores, with disagreement (red as misses, dark blue as false alarms) mostly confined to the boundaries of the eddy, confirming that the network's errors are predominantly boundary-placement errors rather than spurious detections.
Similar conclusions are drawn for NADIRnet, whose results are not shown for simplicity.
The consistency between SWOTnet detections and the ground-truth is further highlighted by the analysis of the induced eddy kinetic energy (EKE) in Fig.~\ref{fig:eke}.
The top panel~\ref{fig:eke_swotnet} shows the global EKE induced by the eddies detected by SWOTnet over the whole training and validation period, while a comparison with the target is drawn in the bottom panel~\ref{fig:eke_diff}.
Here, the zonal averages of the EKE induced both by SWOTnet's and the ground-truth eddies are plotted, so that the similarity between the two is evident.

\begin{figure}[htbp]
    \centering
    \begin{subfigure}{0.48\textwidth}
        \centering
        \includegraphics[width=\linewidth]{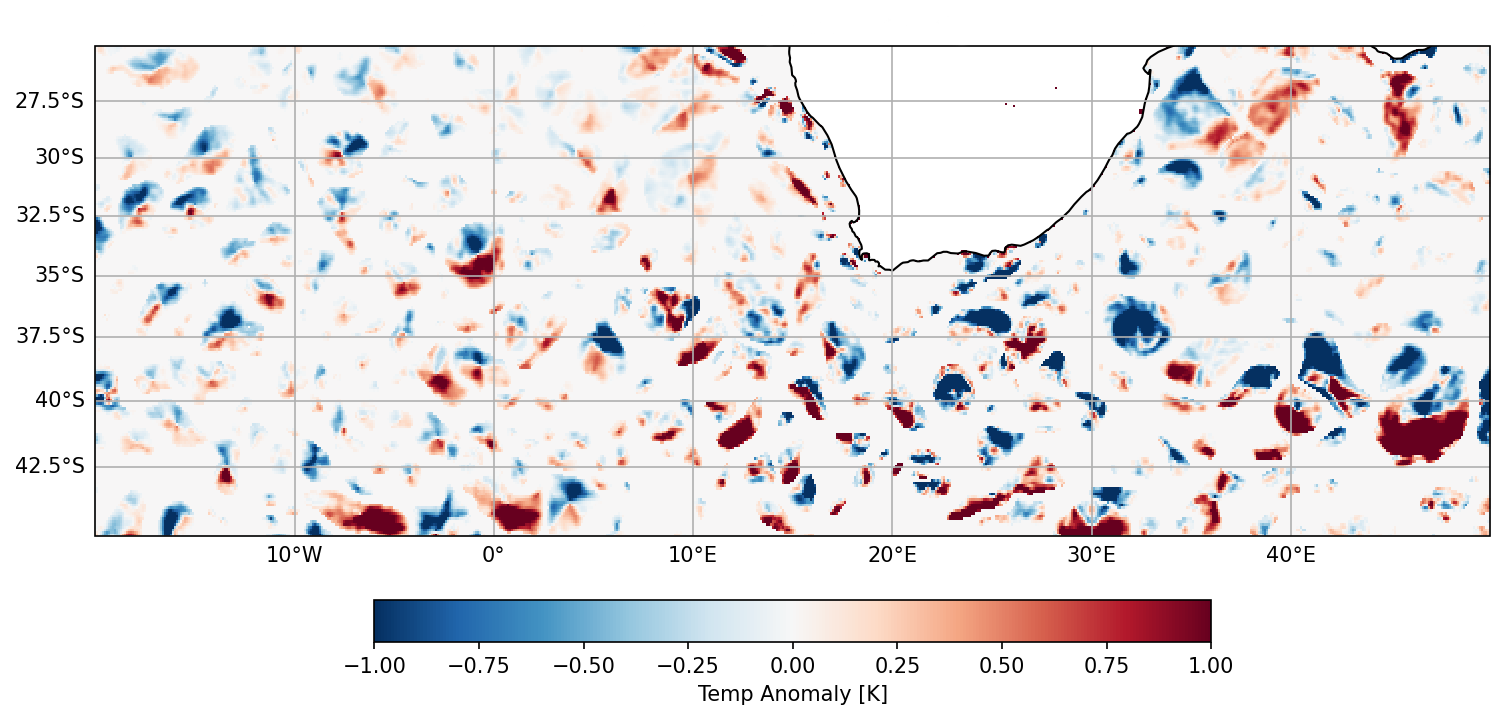}
        \caption{Target (daily)}
        \label{fig:gt_sst}
    \end{subfigure}\hfill
    \begin{subfigure}{0.48\textwidth}
        \centering
        \includegraphics[width=\linewidth]{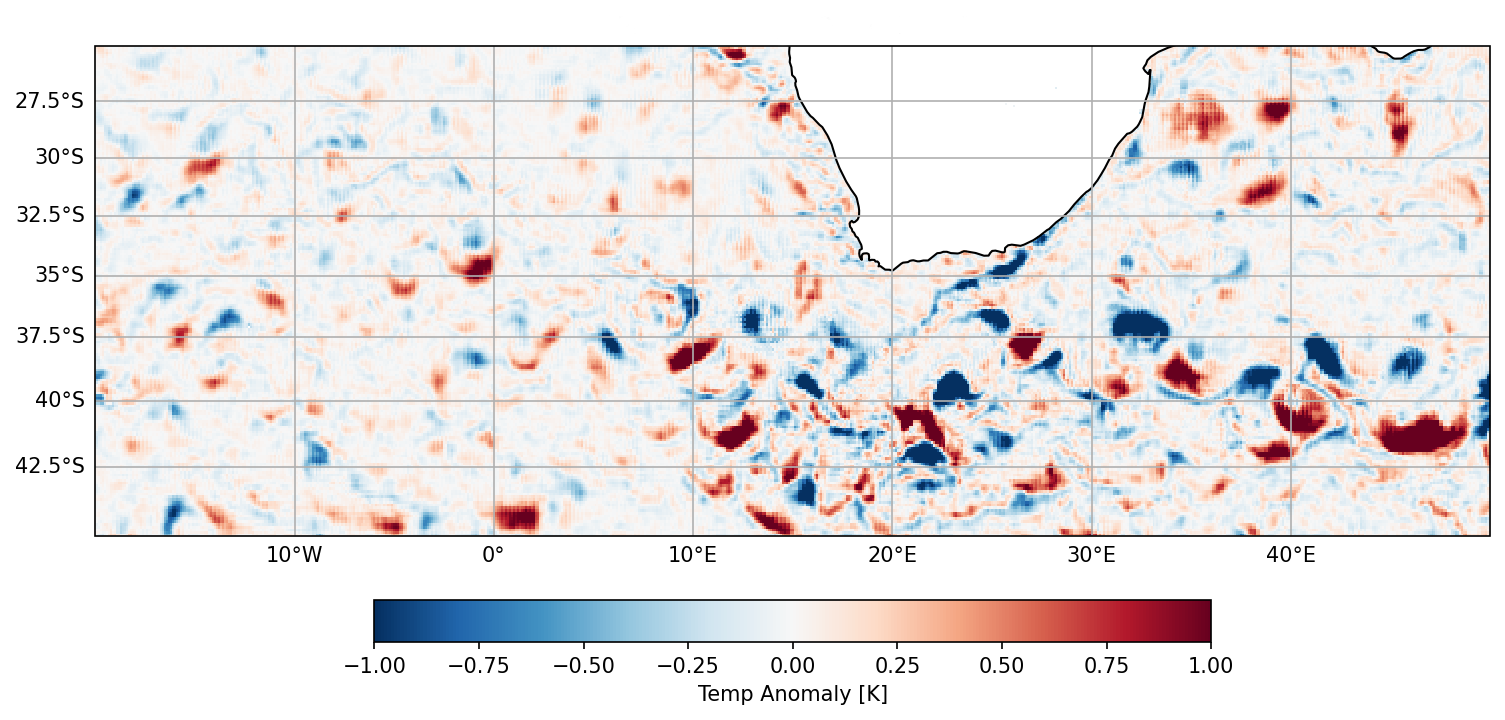}
        \caption{SWOTnet (daily)}
        \label{fig:unet_sst}
    \end{subfigure}

    \vspace{1em} 

    \begin{subfigure}{0.48\textwidth}
        \centering
        \includegraphics[width=\linewidth]{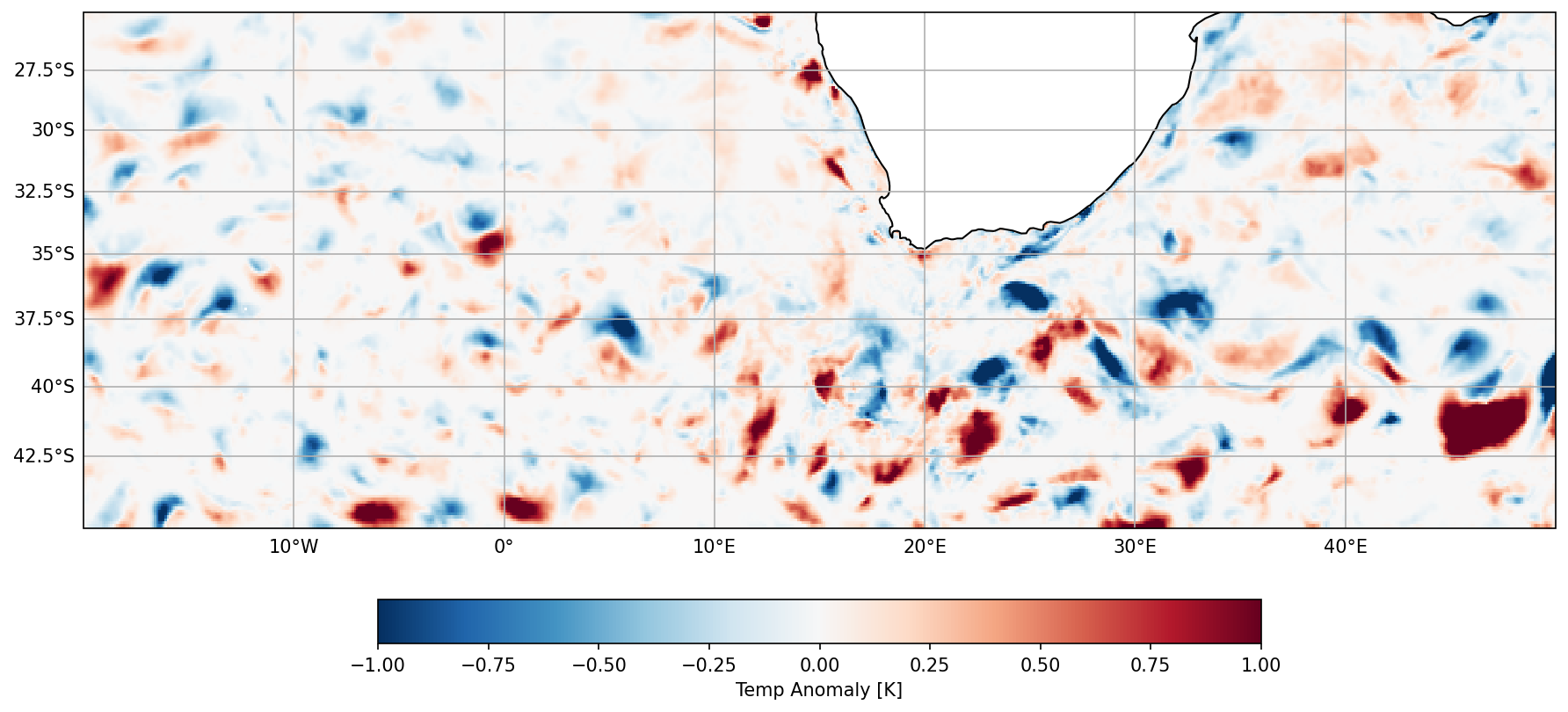}
        \caption{Target (monthly average)}
        \label{fig:gt_sst_monthly}
    \end{subfigure}\hfill
    \begin{subfigure}{0.48\textwidth}
        \centering
        \includegraphics[width=\linewidth]{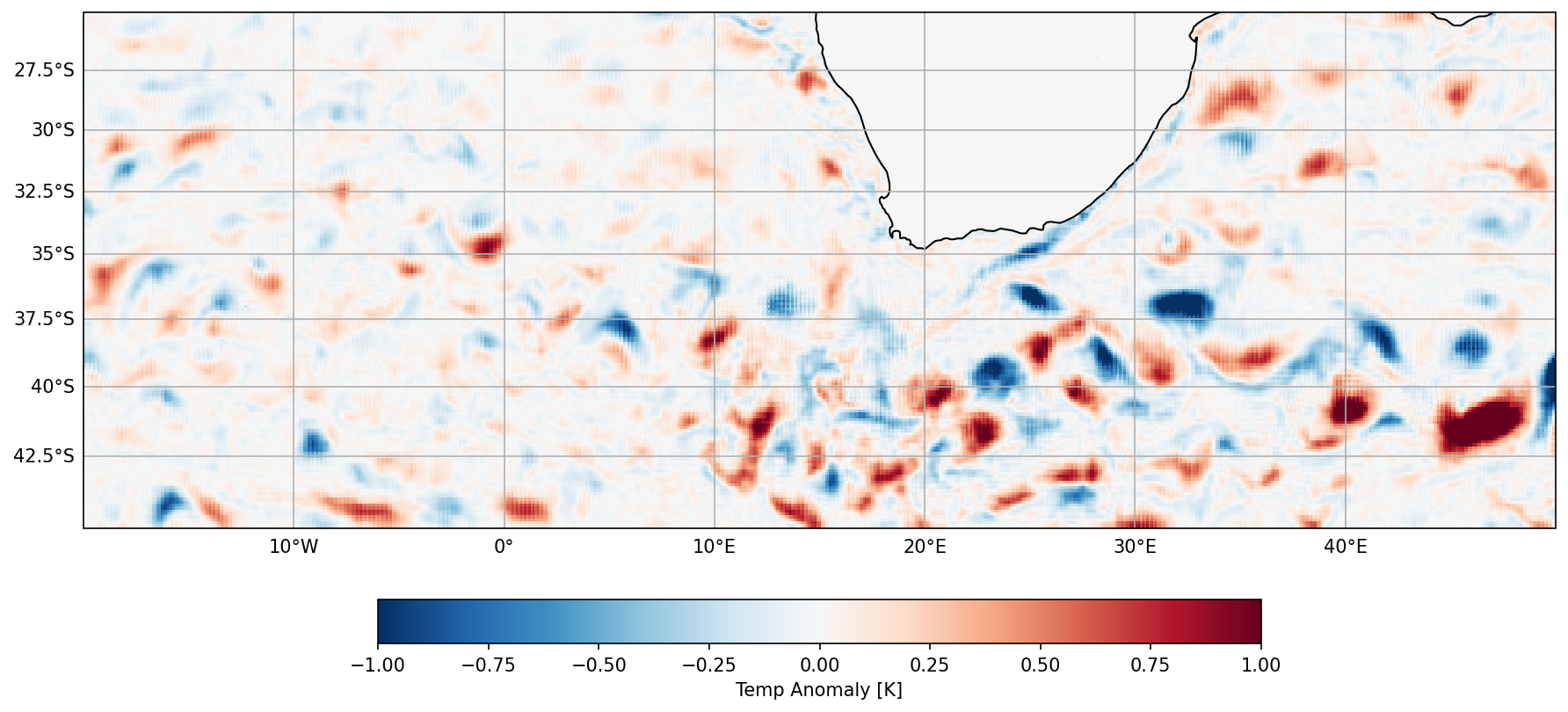}
        \caption{SWOTnet (monthly average)}
        \label{fig:unet_sst_monthly}
    \end{subfigure}
    
    \caption{Spatial SST anomaly reconstructions on 2024-03-27 zoomed-in over the Agulhas region. Top row: daily fields for target~\subref{fig:gt_sst} and SWOTnet~\subref{fig:unet_sst}, showing clear co-location with the eddy cores of Fig.~\ref{fig:eddy_pixels}. Bottom row: 30-day running averages centred on 2024-03-27 for target~\subref{fig:gt_sst_monthly} and SWOTnet~\subref{fig:unet_sst_monthly}, where the broader structural agreement between the two is more evident.}
    \label{fig:sst_output}
\end{figure}

The multi-task nature of the models is illustrated in Fig.~\ref{fig:sst_output}, which compares the target and SWOTnet spatial SST anomaly reconstruction for the same date (top row).
The reconstruction is clearly co-located with the eddy cores identified in Fig.~\ref{fig:eddy_pixels}; the same conclusion holds for NADIRnet (not shown).
When averaging the reconstructions over a month (Fig.~\ref{fig:sst_output}, bottom row), the similarities and broader structural agreement between the two become more evident.

\begin{figure}[htbp]
    \centering
    \begin{subfigure}{0.3\textwidth}
        \centering
        \includegraphics[height=3.5cm]{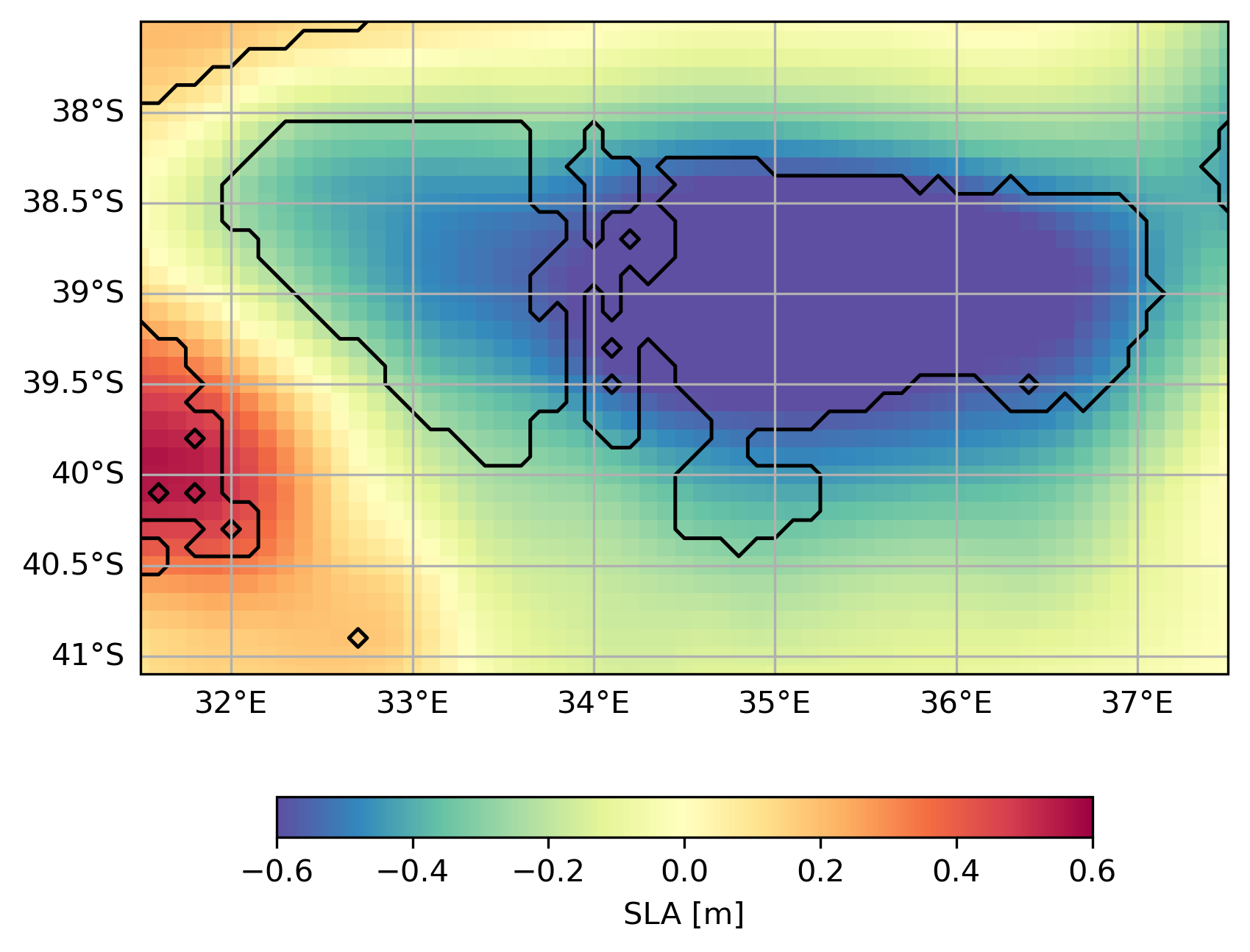}
        \caption{$\mathcal{L} = \mathcal{L}_{BCE}$}
        \label{fig:no_sst}
    \end{subfigure}
    \hfill 
    \begin{subfigure}{0.3\textwidth}
        \centering
        \includegraphics[height=3.5cm]{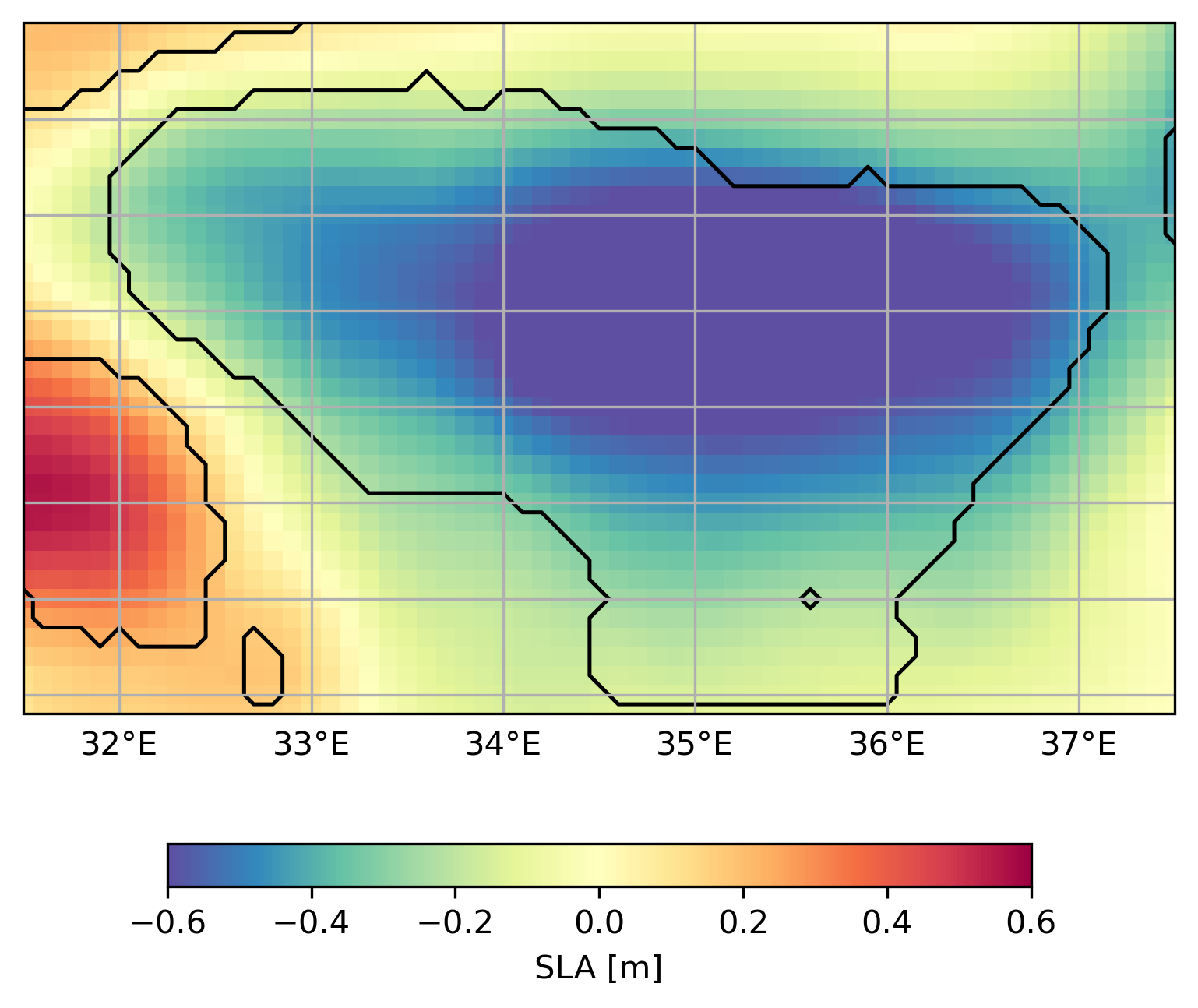}
        \caption{$\mathcal{L} = \mathcal{L}_{total} - \mathcal{L}_{gradient}$}
        \label{fig:boundary_sst}
    \end{subfigure}
    \hfill 
    \begin{subfigure}{0.3\textwidth}
        \centering
        \includegraphics[height=3.5cm]{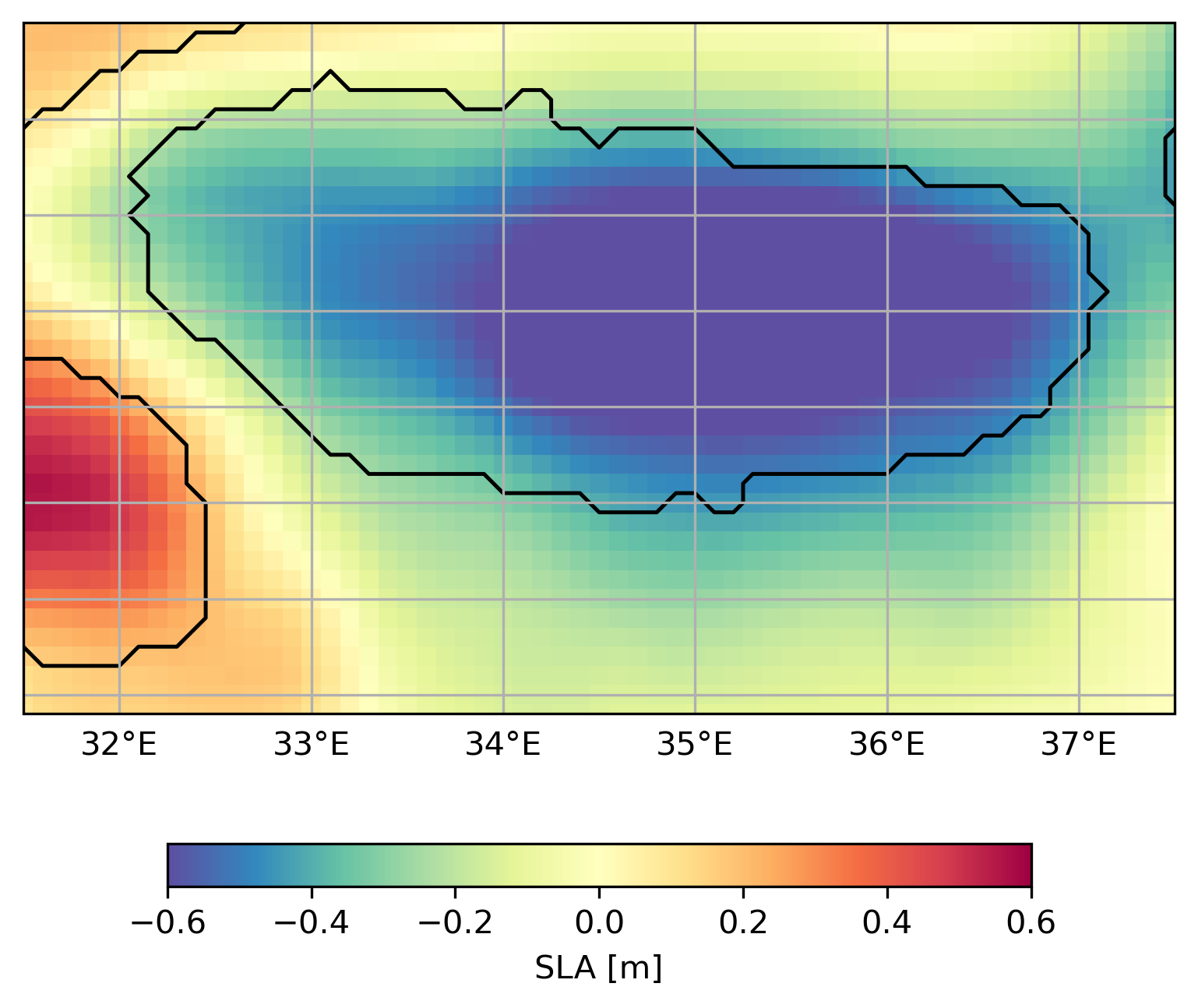}
        \caption{$\mathcal{L} = \mathcal{L}_{total}$}
        \label{fig:full_sst}
    \end{subfigure}

    \caption{Impact of the multi-task training and of the different terms in the loss function on the eddy detection on a selected eddy example. The panels show the SLA field with the ML-based eddy detection contours of different trained models, in particular when the loss function is: \subref{fig:no_sst}~only the detection term and no spatial SST anomaly output channel; \subref{fig:boundary_sst}~the detection term and also the regression and boundary-SST coupling terms with the two-output channels configuration; \subref{fig:full_sst}~as~\subref{fig:boundary_sst} with the addition of the gradient term, i.e.~the total loss function as in eq.~\eqref{eq:total}.}
    \label{fig:sst_impact}
\end{figure}

The contribution of the different loss terms to the sharpness of the detected eddy boundary is illustrated in Fig.~\ref{fig:sst_impact} for a representative eddy.
Training with the detection term alone ($\mathcal{L}=\mathcal{L}_{BCE}$, panel~\ref{fig:no_sst}) might yield an irregular, noisy contour especially when the ML-based detection disagrees with the target, resulting in a structure that does not capture the underlying SLA characteristics.
Introducing the regression and boundary-SST coupling terms (panel~\ref{fig:boundary_sst}) substantially regularises the boundary, which becomes smoother and better aligned with the SLA extremum.
Adding the gradient term (panel~\ref{fig:full_sst}, i.e.~training with the full loss of eq.~\eqref{eq:total}) further sharpens the boundary, producing a contour that closely tracks the SLA gradient.
This progression shows that the auxiliary SST tasks are not only providing additional supervision signals but are consistently improving the geometric quality of the eddy detection itself, supporting the design choice of a coupled, multi-task loss over a detection-only objective.

\begin{figure}
    \centering
    \includegraphics[width=0.75\linewidth]{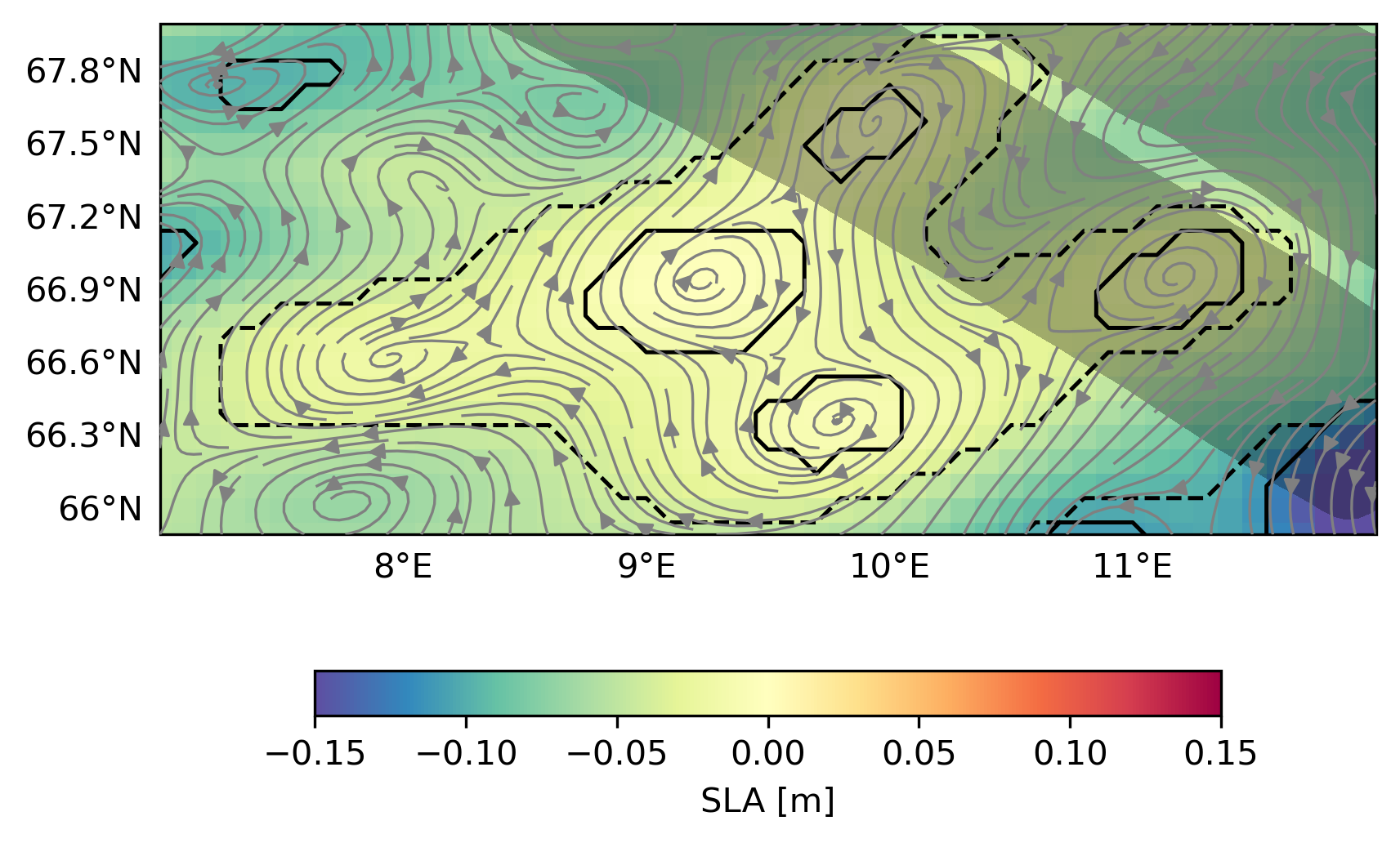}
    \caption{Comparison between the eddy detection of SWOTnet (black solid lines) and the target detection (black dashed lines) on the SLA field with streamlined velocities in the Lofoten Basin of the Norwegian Sea on 2024-03-07; shaded area indicates SWOT coverage on that day. The benefits of the ML-based multimodal detection over the sequential series of constraints of the dynamical one are visible, since the latter could only resolve the local SLA extrema but not the detailed vorticity structures.}
    \label{fig:swotnet_vs_gt_detection}
\end{figure}

The multi-task nature of the ML-based models also helps in regularising their output when additional information can be recovered by the input data but is otherwise imperfect or missing in the target.
This is captured in Fig.~\ref{fig:swotnet_vs_gt_detection}, where SWOTnet's detection (black solid line) is compared with the dynamical detection as target (black dashed line) over an eddy-rich area of the Norwegian Sea, the Lofoten Basin, on 2024-03-07.
Because the dynamical algorithm is based on a sequential series of constraints, whose first is the existence of SLA extrema, it collapses several nearby vorticity structures into a single one.
SWOTnet, by contrast, resolves the finer vorticity structures visible in the streamlined velocity field, recovering individual eddies that the dynamical algorithm merges.
It is worth noting that two of the detected features in Fig.~\ref{fig:swotnet_vs_gt_detection} fall in an area that was covered by SWOT on that day, whose trajectory is indicated by the shaded area, while the remaining two cores were covered by an adjacent pass the following day; this falls within the temporal window of the MIOST algorithm~\parencite{ubelmann_reconstructing_2021}, excluding the possibility that the smaller eddies are only due to interpolation artifacts.
This highlights a key advantage of the multimodal data-driven approach.
Indeed, by considering all input fields at once rather than a sequence of constraints, the network is able to detect structures that a dynamical algorithm might miss by construction, even when the pixel-wise agreement with the target is imperfect.

\begin{figure}
    \centering
    \includegraphics[width=\linewidth]{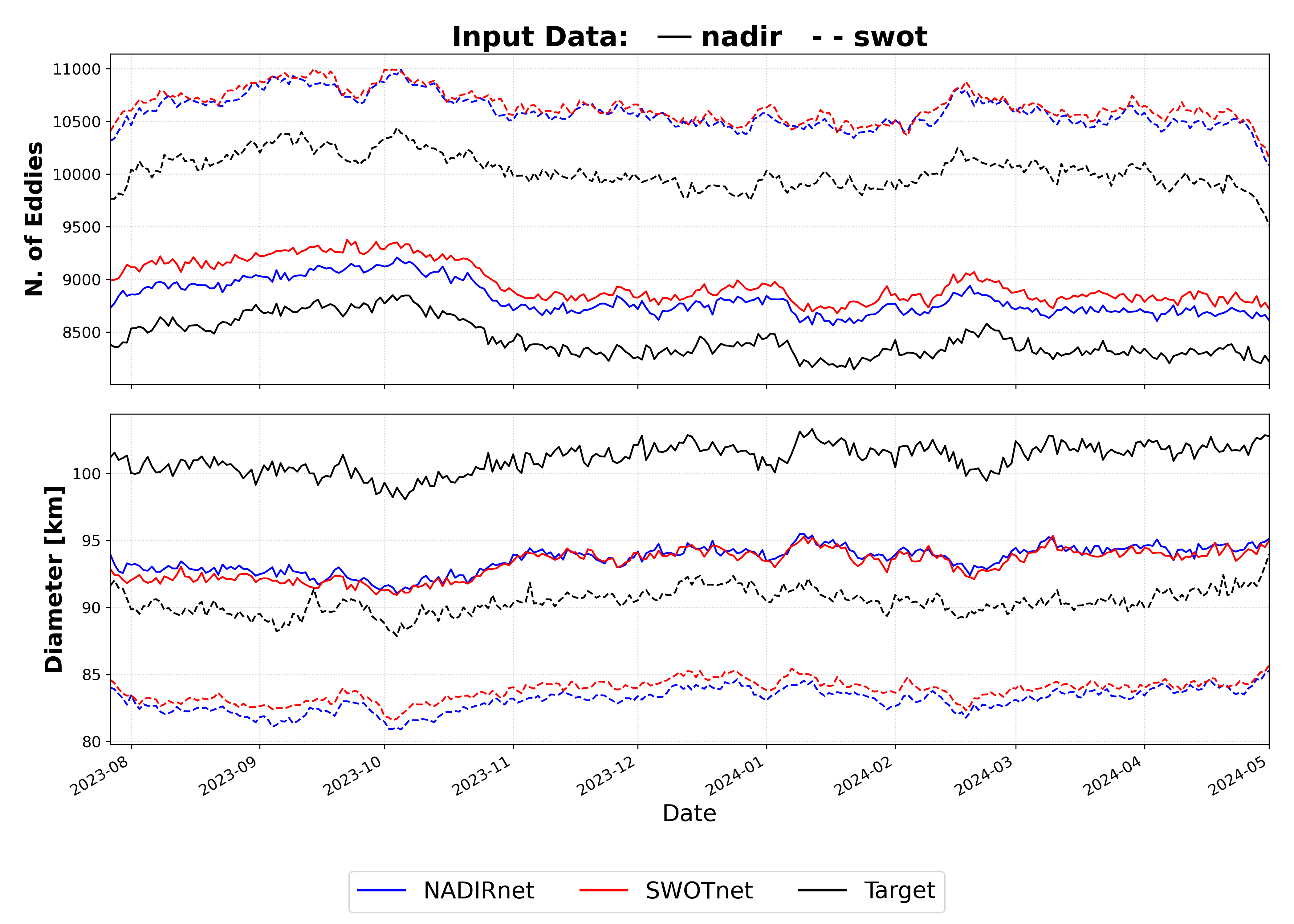}
    \caption{Number (top panel) and average size (bottom panel) of the eddy detected by NADIRnet (blue lines), SWOTnet (red lines) and the dynamical model used as ground-truth (black lines) on different altimetry data, namely conventional nadir (solid lines) and SWOT-enriched (dashed lines) maps, over the training and validation period. Overall, the ML-based models detect a larger population with smaller structures compared to the dynamical algorithm, showing little sensitivity to training data. Nonetheless, SWOTnet is able to recover more details compared to NADIRnet when applied to conventional nadir altimetry, detecting 30\% more eddies compared to the dynamical algorithm.}
    \label{fig:timeseries}
\end{figure}

An analysis of the population statistics over the training and validation period, presented in Fig.~\ref{fig:timeseries}, suggests that this behaviour is systematic.
Both ML-models consistently detect a larger population of eddies (daily, about 500 structures more) with a size of about $7\,\text{km}$ smaller in diameter on average compared to the target detection.
This is true both for nadir and SWOT-derived altimetry, with little sensitivity to which altimetry product is considered as input, stressing the robustness of the ML models.
Nonetheless, SWOTnet notably recovers more structures than NADIRnet even when applied to conventional nadir altimetry, detecting approximately 30\% more eddies than the dynamical algorithm on this lower-resolution product.
This suggests that the higher-resolution training signal, thanks to the SST anchor, transfers to coarser input data, supporting the use of a single SWOT-trained model as a candidate for reprocessing the full multi-decadal nadir altimetry record.

\subsection{Applications}

\begin{figure}
    \centering
    \includegraphics[width=0.5\linewidth]{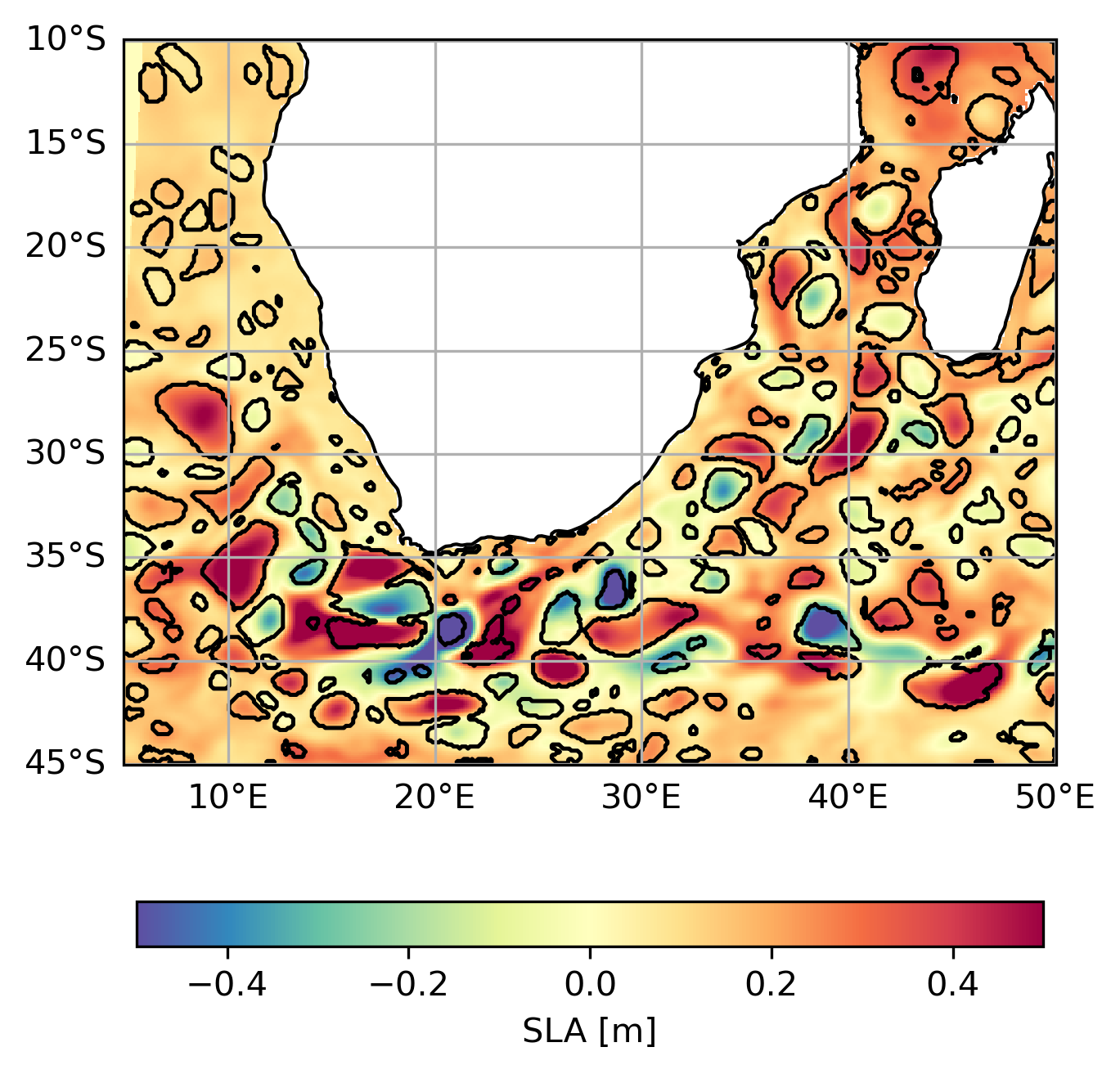}
    \caption{Example of eddy detection with SWOTnet on the VarDyn product produced in the Agulhas region for 2024-06-01.}
    \label{fig:vardyn_detection}
\end{figure}

\begin{figure}[htbp]
    \centering
    \begin{subfigure}{0.9\textwidth}
        \centering
        \includegraphics[width=\linewidth]{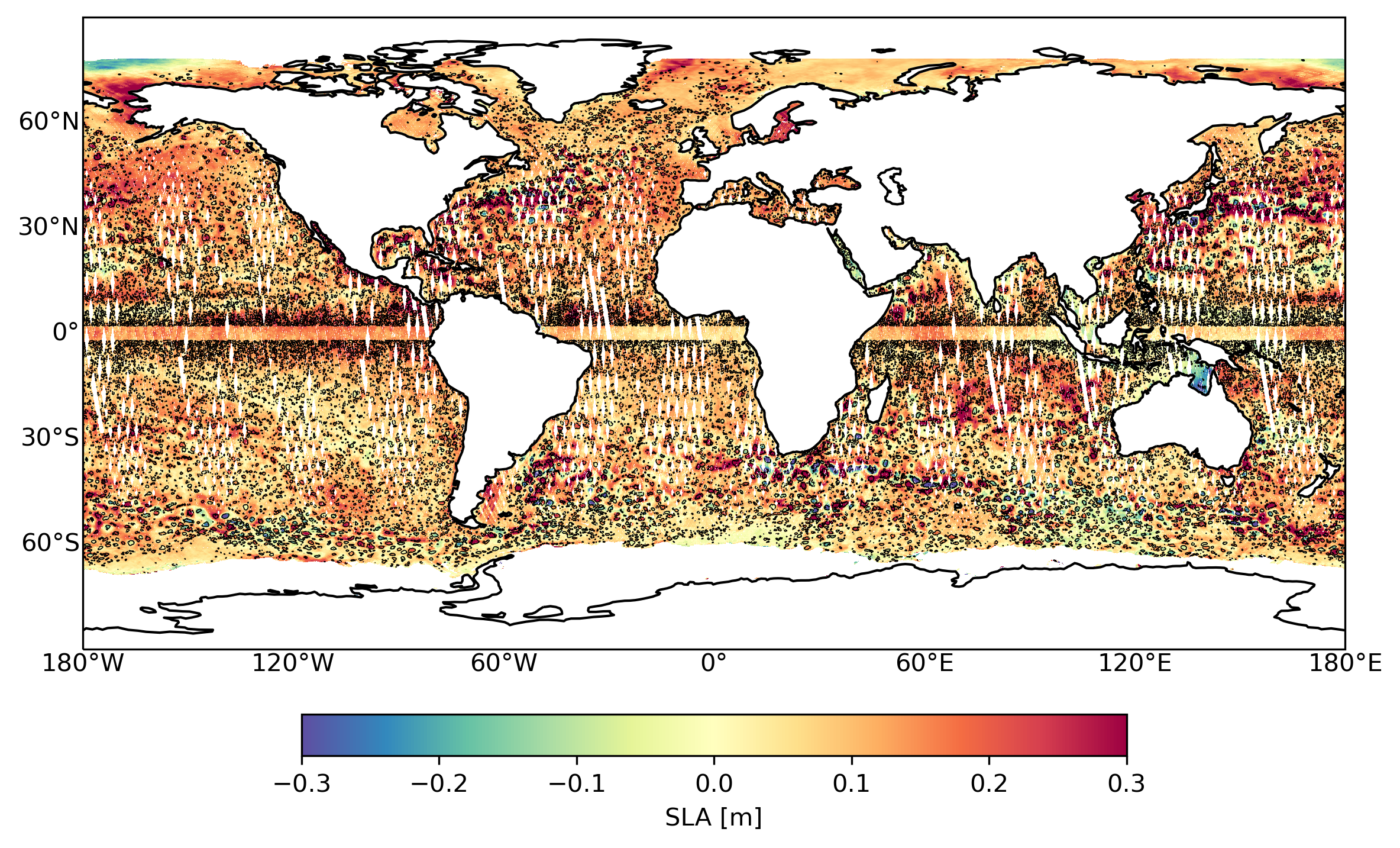}
        \caption{Global domain}
        \label{fig:l3_global}
    \end{subfigure}
    \vspace{1em} 
    \begin{subfigure}{0.5\textwidth}
        \centering
        \includegraphics[width=\linewidth]{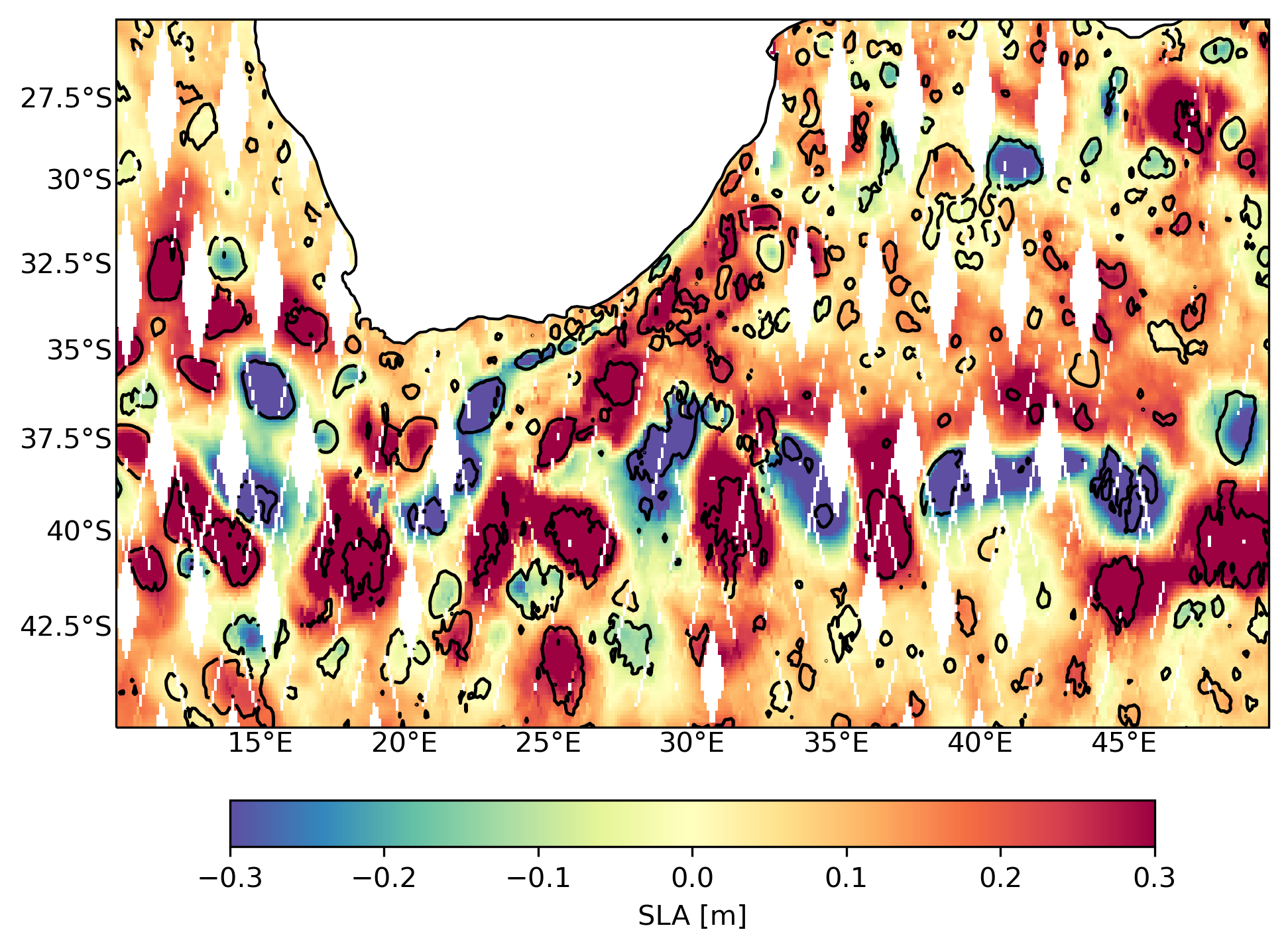}
        \caption{Agulhas region}
        \label{fig:l3_agulhas}
    \end{subfigure}

    \caption{Composite images of eddy detection on SWOT L3 data with SWOTnet. The detection is performed on the available daily global swathes of a single cycle, which are then superimposed together to create the composite image~\subref{fig:l3_global}, which is also shown zoomed-in over the Agulhas region \subref{fig:l3_agulhas}.}
    \label{fig:l3}
\end{figure}

The robustness of the trained models to products not seen during training is illustrated in Figs.~\ref{fig:vardyn_detection} and~\ref{fig:l3}.
Figure~\ref{fig:vardyn_detection} shows SWOTnet detection on the VarDyn product~\parencite{le_guillou_vardyn_2025} in the Agulhas region.
VarDyn jointly reconstructs physically consistent SSH and SST maps by assimilating satellite altimetry and SST observations into a 1.5-layer quasi-geostrophic model coupled to an SST advection–diffusion model within a variational framework, with controlled error fluxes accounting for model imperfections.
The Agulhas product used here provides daily maps on a $0.1^\circ \times 0.1^\circ$ grid.
Its combination of observational information and dynamical constraints provides a complementary test of SWOTnet’s ability to generalise beyond the mapping approaches represented in its training data.
When the network is applied to VarDyn's altimetry and SST fields, it produces coherent, physically plausible eddy contours that closely track the SLA extrema of the VarDyn field, indicating that the model has learned the physical condition leading to a detection generalising across mapping methodologies, rather than overfit to the specific characteristics of the training products.

As an example of real-world applicability of SWOTnet, we note that the whole detection process on a single day of the VarDyn product over the region in Fig.~\ref{fig:vardyn_detection} can run on the CPU of a personal laptop in less than a minute taking about $3\,\textrm{GB}$ of memory. The process consists of pre-processing of the data to make it available to the network, running the network to detect eddies and reconstruct the spatial SST anomaly field, post-processing the output to identify thermodynamically active eddies and writing the output to file.
This enabled the network to be tested on board the Marion Dufresne vessel during the field campaign of the WHIRLS project\footnote{\url{https://www.whirls.eu/}.} in July 2026, providing daily information on the eddy positions to support the cruise planning in addition to the operational ToEddies algorithm~\parencite{laxenaire_toeddies_2026}. 

A further application case is shown in Fig.~\ref{fig:l3}, where SWOTnet is applied to SWOT L3 wide-swath retrievals instead of the gap-filled L4 maps employed for training.
The detection is performed daily on the globally-available swathes for a sample SWOT cycle, and then composited in time to create a denser picture.
Figure~\ref{fig:l3} shows the resulting image for the global domain, and also zoomed-in over the Agulhas region.
The resulting composites are spatially coherent across neighbouring and overlapping swathes, despite the model never having been trained on gapped data.
This, besides confirming that the model's detection is physics- and feature-based, suggests that the model could be considered for near-real-time applications on raw SWOT L3 data.

We note, however, a practical caveat when applying the ML-based models to datasets that differ from those used for training and validation.
The operating thresholds reported in Tab.~\ref{tab:detection_scores} were selected to maximise F\textsubscript{1} score on the DUACS and MIOST validation products.
Those thresholds correspond, after the sigmoid activation, to eddy-probability of $\sigma(0.417)\approx0.60$ and
$\sigma(0.208)\approx0.55$.
These values are not guaranteed to remain optimal when the model is applied to a product with different resolution, noise level or processing chain, such as VarDyn or the L3 composites shown above.
For instance, for the L3 detection shown above a different threshold has been chosen, corresponding to an eddy-probability $\sigma(2)\approx0.88$ in order to avoid noisy detection at the swathes borders.

The choice of threshold should be guided by which type of error is more costly for the intended application.
Raising the probability threshold above the validation-optimised value makes the model more conservative, reducing false positives at the price of missing more uncertain eddies; lowering the threshold has the opposite effect, recovering more structures but admitting more spurious detections.

\section{Conclusion}\label{sec:conclusions}
We presented a multimodal, multi-task, data-driven eddy detection framework based on a shallow U-Net architecture, jointly trained to segment mesoscale eddies and reconstruct their associated spatial SST anomaly.
Two model variants were developed and compared: NADIRnet, trained on conventional nadir altimetry, and SWOTnet, trained through a two-phase curriculum learning strategy that pre-trains on conventional nadir maps and then fine-tunes on SWOT-enriched maps.
Both models achieve high detection scores against the target at pixel and structure level, with the majority of the residual disagreement attributable to boundary placement rather than to missed or spurious detections, as shown by the structure-wise outcome statistics and by direct visual inspection, as well as an analysis of the EKE induced by the eddies.

The multi-task loss, in particular the boundary-SST coupling and gradient-matching terms, was shown to meaningfully improve the geometric quality of the detected eddy boundaries beyond what the detection term alone provides, confirming that the auxiliary SST regression task acts as an effective physical regulariser.
Compared against a conventional dynamically-based detection, the proposed models consistently recover a larger eddy population with on average smaller sizes.
This benefit is not confined to the higher-resolution SWOT-enriched data.
Indeed, SWOTnet's higher-resolution training signal transfers to conventional nadir altimetry, where it detects approximately 30\% more eddies than the dynamical algorithm and more structures than NADIRnet itself, supporting the use of a single SWOT-trained model to reprocess the full multi-decadal nadir altimetry record.

The framework was further shown to generalise well beyond the specific products and processing levels seen during training.
Applied without any retraining to the VarDyn product and to SWOT wide-swath L3 retrievals, the model produced coherent, physically plausible detections, including across gaps in the data, suggesting that the network has learned local, feature-based detection criteria rather than overfitting to the noise or interpolation characteristics of the training products.
Combined with its low computational cost --- a full detection cycle for a single day runs in under a minute on a laptop CPU --- this robustness makes the network a good candidate for operational deployment.

We also highlighted a practical caveat for users applying these models to new datasets. The operating logit threshold reported here was tuned to maximise F\textsubscript{1} score on the specific validation products used in this study, and its optimal value is expected to shift with the resolution, noise characteristics, or processing chain of a new product, as illustrated by the higher threshold adopted for the L3 wide-swath composites.
It is thus recommended to choose the threshold according to the relative cost of false positives versus false negatives for the desired application, rather than assuming the validation-derived value transfers unchanged.

Future work will explore reprocessing the full satellite altimetry record since 1993 with the SWOTnet trained model, with the goal of providing a higher-resolution mesoscale eddy reconstruction across the whole altimetry era.
An extensive analysis of the results on gapped data would also be desirable, in order to assess whether a fine-tuning on, for instance, L3 altimetry is beneficial for near-real-time application.
Furthermore, building on the multimodal nature and robustness of the network to input data, it is also possible to employ the network as a downstream validation tool to assess the consistency of altimetry and temperature maps, both from observation or model data.
Model weights, training and evaluation code are publicly accessible~\parencite{broccoli_water-ml-eddy-detection_2026}, so that the network is available for further studies or applications.

\section*{Acknowledgments}
This study has been conducted using~\cite{cmems_conventional_altimetry}.
The VarDyn data used in this study are available upon reasonable requests to the authors.
The work that led to this manuscript was carried out under a programme of, and funded by, the European Space Agency through the Wide-swath Altimetry for Eddy Reconstruction (WATER) Project; the view expressed here can in no way be taken to reflect the official opinion of the European Space Agency.

\printbibliography

\end{document}